\documentclass[preprint,12pt]{elsarticle}

\usepackage{amssymb}
\usepackage{amsmath}

\usepackage{booktabs}

\journal{Physica A}

\begin{document}

\begin{frontmatter}



\title{Trinity of Varentropy: \\Finiteness, Fluctuations, and Stability \\in Power-Law Statistics} 


\author{Hiroki Suyari} 

\affiliation{organization={Graduate School of Informatics, Chiba University},
            addressline={1-33, Yayoi-cho, Inage-ku,}, 
            city={Chiba},
            postcode={263-8522}, 
            state={Chiba},
            country={Japan}}

\begin{abstract}
Power-law distributions are widely observed in complex systems, yet establishing their thermodynamic consistency remains a theoretical challenge. In this paper, we present a thermodynamic framework for power-law statistics based on the \textit{renormalized entropy} $s_{2-q}$. Derived from the asymptotic scaling of the combinatorial $q$-factorial, this quantity yields a stable thermodynamic limit, remaining finite ($O(N^0)$) for systems with strong correlations. Furthermore, we clarify the physical origin of the nonlinearity parameter $q$ through the concept of \textit{Varentropy} (Variance of Entropy). By mapping the microscopic combinatorics sequentially onto the probability constraints of Type-B superstatistics, we prove that the entropic index is strictly governed by the exact identity $q-1 = 1/C$, where $C$ is the total thermodynamic heat capacity of the system and the reservoir. This deductive derivation, enforced by the uniqueness of the inverse Laplace transform, demonstrates that power-law statistics emerges as a necessary consequence of finite thermal environments, providing a first-principles description beyond the standard infinite Boltzmann-Gibbs limit ($C \to \infty$).
\end{abstract}








\end{frontmatter}




\section{INTRODUCTION}
\label{sec:introduction}

Power-law distributions are a common feature of complex systems.
From high-energy particle collisions~\cite{Wilk2000} and turbulent flows~\cite{Beck2001} to financial markets~\cite{Borland2002} and biological networks, systems with strong correlations or long-range interactions consistently exhibit heavy-tailed statistics that deviate from the standard Boltzmann-Gibbs description.
To capture these phenomena, non-extensive statistical mechanics, pioneered by Tsallis~\cite{Tsallis1988}, has provided a framework, generalizing the standard exponential factor to the $q$-exponential function.
Despite its empirical success, a theoretical challenge remains: the consistent definition of the \textit{thermodynamic limit}.
In standard statistical mechanics, the extensivity of entropy (i.e., entropy proportional to system size $N$) is necessary for the existence of stable thermodynamic potentials.
However, for systems governed by power-law statistics, the standard non-extensive entropy $S_q$ generally becomes non-additive and loses the extensive scaling property ($S_q \propto N^{2-q}$ or similar anomalous scaling) as the system size diverges~\cite{Tsallis2009}.
This scaling behavior has raised questions regarding the physical validity of non-extensive thermodynamics, particularly when applied to classical Hamiltonian systems \cite{Lutsko2011}.
Consequently, a fundamental debate has emerged: Does a consistent thermodynamics exist for finite systems that exhibit power-law behavior?

In this paper, we address this issue by establishing a thermodynamic framework mathematically grounded in the combinatorics of the $q$-factorial and physically interpreted through the concept of \textit{Varentropy} (Variance of Entropy).
We show that the anomalous scaling is a consequence of the finite-size effects inherent in correlated systems.
Our approach introduces a generalized thermodynamic state variable, the \textit{renormalized entropy} $s_{2-q}$, which is constructed to remain finite ($O(N^0)$) in the thermodynamic limit, thereby resolving the stability problem.

Furthermore, we investigate the physical origin of the non-extensivity parameter $q$.
While $q$ is often treated as a fitting parameter, we show that it is a direct measure of the \textit{finite heat capacity} of the environment.
By linking the macroscopic variational principle with the microscopic theory of Superstatistics~\cite{BeckCohen03}, we derive a relation linking $q$ to the heat capacity $C$ of the reservoir: $|q-1| \simeq 1/C$.
This result implies that power-law statistics describes the thermodynamics of finite systems, where the fluctuations of intensive parameters (temperature) cannot be neglected.

Unlike typical approaches in superstatistics that assume a specific temperature fluctuation distribution $f(\beta)$ a priori, the present framework adopts a deductive construction. The formulation originates from a single microscopic differential equation governing the scale-invariant expansion of the phase volume, $\frac{dW(N)}{dN} \propto W(N)^q$. From this foundational equation, the complete macroscopic thermodynamic structure is sequentially deduced. First, the differential equation generates the fundamental structure of Varentropy. Second, by incorporating the framework of Type-B superstatistics to ensure proper probabilistic consistency as initiated by Sattin \cite{Sattin2006}, we demonstrate that the functional form of the macroscopic temperature fluctuation $f(\beta)$ is uniquely determined. Due to the uniqueness of the inverse Laplace transform, $f(\beta)$ necessarily takes the form of a Gamma distribution. This deductive sequence demonstrates that the $q$-canonical distribution and Type-B superstatistics are self-consistent macroscopic manifestations derived from the same microscopic combinatorial origin, thereby providing a first-principles foundation rather than a phenomenological postulate.

The paper is organized as follows.
In Sec.~\ref{sec:theoretical_framework}, we establish the mathematical foundations, deriving the $q$-factorial growth rule and introducing the renormalized entropy $s_{2-q}$.
In Sec.~\ref{sec:physical_interpretation}, we explore the physical meaning of $q$ through the expansion of entropy and the concept of Varentropy.
In Sec.~\ref{sec:micro_foundations}, we construct the microscopic foundations using Type-B superstatistics, demonstrating that the functional form of temperature fluctuations is uniquely restricted to a Gamma distribution via the inverse Laplace transform.
In Sec.~\ref{sec:macroscopic_consequences}, we deduce the macroscopic consequences of these fluctuations, establishing the exact identity $q-1=1/C$ and its direct relation to thermodynamic stability and anomalous scaling.
Finally, Sec.~\ref{sec:conclusion} summarizes our findings and discusses the implications for finite-size thermodynamics.


\section{THEORETICAL FRAMEWORK}
\label{sec:theoretical_framework}

In this section, we outline the mathematical foundations for power-law statistics.
Rather than postulating a macroscopic entropy functional a priori, we analyze how the number of microstates grows with the system size. A nonlinear generalization of the growth rule leads to the $q$-algebra and the associated combinatorial structure.

\subsection{Generalized Phase Space Growth}
\label{subsec:Generalized Phase Space Growth}

Consider a physical system with an effective phase space volume $W(N)$ dependent on the system size $N$.
For standard extensive systems with short-range interactions, the phase space volume grows exponentially, governed by the linear differential equation $dW/dN = \lambda W$, where $\lambda$ is a growth constant.
However, for systems characterized by long-range correlations or strong entanglement, the growth rate may depend non-linearly on the current volume.
We generalize the growth law as:
\begin{equation}
\frac{dW}{dN} = \lambda W^q,
\label{eq:diff_eq}
\end{equation}
where the exponent $q$ characterizes the degree of non-extensivity and the nature of the phase space correlations.
To ensure thermodynamic stability, specifically the concavity of the entropy functionals $S_q$ and $S_{2-q}$, the physical regime is strictly bounded to $0 < q < 2$ throughout this paper. Within this range, the system exhibits distinct growth regimes:
\begin{itemize}
    \item $q=1$: Standard extensive systems (Independent growth).
    \item $0 < q < 1$: Restricted growth (Phase space is suppressed by correlations).
    \item $1 < q < 2$: Super-extensive growth (Phase space is expanded).
\end{itemize}

Here, the direct connection between the entropy and the phase-space volume strictly requires the assumption of equal \textit{a priori} probabilities, which is fundamentally applicable to both the standard Boltzmann entropy ($q=1$) and the $S_q$ entropy.

To analyze the solution $W(N)$, it is useful to consider the transformation that linearizes the growth.
Rearranging Eq.~(\ref{eq:diff_eq}) as $W^{-q} dW/dN = \lambda$, we can express this non-linear growth equivalently using the \textit{$q$-logarithm} ($\ln_q x := \frac{x^{1-q}-1}{1-q}$):
\begin{equation}
\frac{d \ln_q W}{dN} = \lambda.
\label{eq:q_log_diff}
\end{equation}
Equation (\ref{eq:q_log_diff}) shows that the $q$-logarithm transforms the power-law growth into a constant intensive rate.
Integrating this equation yields the linear relation:
\begin{equation}
\ln_q W = \lambda N + \text{const}.
\label{eq:q_log_linear}
\end{equation}
The $q$-logarithm serves as a natural scaling function that maps the correlated phase space volume $W$ into a linear extensive quantity.
To maintain the extensivity of the macroscopic state variable, the entropy of the system is defined as:
\begin{equation}
S_q := k_B \ln_q W.
\label{eq:tsallis_entropy_def}
\end{equation}
While the inverse relation is given by the $q$-exponential function ($W = \exp_q(\lambda N + \text{const})$), the linearity of Eq.~(\ref{eq:q_log_linear}) underpins the generalized thermodynamics.

It should be noted that the connection between anomalous phase-space growth and extensive non-standard entropies has been formally explored from various perspectives, such as the comprehensive classification of complex systems based on asymptotic growth rates \cite{Hanel2011} and formal group approaches \cite{Tempesta2011}.

\subsection{Algebraic Structure: Linearization via $q$-Logarithm}
The nonlinear growth rule Eq.~(\ref{eq:diff_eq}) implies that the composition of subsystems is no longer multiplicative.
If we consider a composite system formed by two subsystems $A$ and $B$, the total number of microstates $W_{A \cup B}$ involves a generalized product $W_A \otimes_q W_B$ due to the presence of correlations.
We define the composition rule $\otimes_q$ by the requirement that the generalized entropies are additive:
\begin{equation}
\ln_q (W_A \otimes_q W_B) = \ln_q W_A + \ln_q W_B.
\label{eq:q_additivity}
\end{equation}
This relation asserts that, although the microstate volumes $W$ interact nonlinearly via the generalized $q$-product \cite{Nivanen2003,Borges2004}, the corresponding variables $\ln_q W$ follow the standard addition rule.
Thus, the $q$-logarithm transforms the nonlinear composition into a linear superposition, serving as the appropriate variable for describing correlated systems.

\subsection{$q$-Stirling's Formula}

By extending the generalized $q$-product to a sequence of successive integers, the generalized $q$-factorial $N!_q$ is defined as:
\begin{equation}
N!_q := 1 \otimes_q 2 \otimes_q \cdots \otimes_q N.
\label{eq:q_factorial_def}
\end{equation}
Taking the $q$-logarithm of both sides yields $\ln_q (N!_q) = \sum_{i=1}^N \ln_q i$. For large $N$, the direct analytical evaluation of this generalized factorial is intractable. We therefore employ the $q$-deformation of Stirling's approximation \cite{Suyari2006}, hereafter referred to as the \textit{$q$-Stirling's formula}. Consistent with the classification presented in Section \ref{subsec:Generalized Phase Space Growth}, this approximation is applicable across the growth regimes: restricted growth ($0 < q < 1$), standard extensive growth ($q=1$), and super-extensive growth ($1 < q < 2$). Assuming the parameter range $0 < q < 2$, the $q$-Stirling's formula is given by:
\begin{equation}
\ln_q (N!_q) \simeq N \ln_q N - N.
\label{eq:q_stirling}
\end{equation}
This expression reduces to the standard Stirling's formula, $\ln (N!) \simeq N \ln N - N$, in the limit $q \to 1$. For $q \neq 1$, Eq.~(\ref{eq:q_stirling}) captures the non-linear scaling of the phase space volume, providing a tool for deriving the macroscopic thermodynamic properties of the reservoir.

\subsection{$q$-Multinomial Coefficient and Entropy}
We define the \textit{$q$-multinomial coefficient} as the effective number of microstates for a system partitioned into $k$ groups with sizes $N_1, \dots, N_k$ ($N = \sum N_i$).
Using the $q$-logarithm and the generalized additivity, this coefficient is given by:
\begin{equation}
\ln_q \left( \begin{array}{c} N \\ N_1 \dots N_k \end{array} \right)_q := \ln_q N!_q - \sum_{i=1}^k \ln_q N_i!_q.
\label{eq:q_multinomial_def}
\end{equation}
This quantity generalizes the counting factor to correlated systems.
Substituting the generalized Stirling's formula (Eq.~\ref{eq:q_stirling}) into Eq.~(\ref{eq:q_multinomial_def}), we obtain the asymptotic relation linking the microstate count to the macroscopic entropy:
\begin{equation}
\ln_q \left( \begin{array}{c} N \\ N_1 \dots N_k \end{array} \right)_q \simeq \frac{N^{2-q}}{2-q} S_{2-q}^{\text{Tsallis}}(p_1, \dots, p_k),
\label{eq:entropy_relation}
\end{equation}
where $S_{q}^{\text{Tsallis}}(p) := \frac{1 - \sum p_i^q}{q-1}$ is the Tsallis entropy.
This establishes a mathematical link \cite{Suyari2006}: the combinatorial structure
 of the $q$-multinomial coefficient corresponds to the Tsallis entropy with index $2-q$ in the thermodynamic limit.

The mathematical scaling of $N^{2-q}$ and the associated dual relation between the indices $q$ and $2-q$ via the $q$-generalized multinomial coefficient were first rigorously established by the author in \cite{Suyari2006}. While the existence of this mathematical duality has been acknowledged in subsequent literature \cite{Tsallis2009}, the primary contribution of the present work is to advance this relation from a formal combinatorial identity into a physically grounded, intensive thermodynamic state variable that secures macroscopic equilibrium.

\subsection{The Generalized Thermodynamic Limit}

An issue in establishing a statistical mechanics for power-law systems is the thermodynamic limit.
As a direct thermodynamic consequence of the combinatorial relation obtained in Eq.~(\ref{eq:entropy_relation}), the total non-extensive entropy derived from the $q$-multinomial coefficient, defined as $S_q(N) := \ln_q \left( \begin{array}{c} N \\ N_1 \dots N_k \end{array} \right)_q$, scales asymptotically as:
\begin{equation}
S_q(N) \sim N^{2-q}.
\end{equation}
By dividing both sides of this relation by the growth factor $N^{2-q}$, the renormalized, intensive entropy $s_{2-q}$ is naturally defined as:
\begin{equation}
s_{2-q} := \lim_{N \to \infty} \frac{S_q(N)}{N^{2-q}}.
\label{eq:intensive_entropy_def}
\end{equation}
This formulation clearly demonstrates that the $N^{2-q}$ rescaling is not an ad-hoc adjustment, but a mathematically required operation grounded in the microscopic configurations of the system to yield a finite and stable thermodynamic limit.
Table \ref{tab:thermo_vars} summarizes the distinction between the unscaled total entropy and the renormalized intensive state variable.

\begin{table}[h]
\centering 
\caption{\label{tab:thermo_vars}Definition of thermodynamic variables in power-law statistics.
Note that the renormalized entropy $s_{2-q}$ serves as the intensive state variable governing equilibrium.}
\begin{tabular}{lcc}
\toprule 
Quantity & Scaling ($N \to \infty$) & Physical Role\\
\midrule 
Total Entropy $S_q(N)$ & $\sim N^{2-q}$ & \begin{tabular}{@{}c@{}}Measure of \\ microstates\end{tabular} \\
Renormalized Entropy $s_{2-q}$ & $\sim N^0$ (Finite) & \begin{tabular}{@{}c@{}}Intensive \\ state variable\end{tabular} \\
\bottomrule 
\end{tabular}
\end{table}

By constructing thermodynamics based on $s_{2-q}$, we ensure a stable macroscopic description.
The expansion of this renormalized entropy $s_{2-q}$ around $q=1$ relates to the physical origin of the parameter $q$, which will be the focus of the next section.

\subsection{Variational Principle and the $q$-Canonical Distribution}
We determine the equilibrium probability distribution by applying the Maximum Entropy principle to the renormalized entropy established above.
A feature of the non-extensive formalism is the duality between the entropy index and the resulting distribution index when standard constraints are employed.
As shown by Wada and Scarfone~\cite{Wada2005}, maximizing the entropy with index $2-q$ (our renormalized entropy $s_{2-q}$) under the linear mean energy constraint $U = \sum_{i} p_i E_i$ yields the $q$-exponential distribution.
The variational functional is given by:
\begin{equation}
\frac{\delta}{\delta p_i} \left( s_{2-q} - \beta \sum_{j} p_j E_j - \alpha \sum_{j} p_j \right) = 0,
\end{equation}
where $\beta$ and $\alpha$ are Lagrange multipliers associated with the internal energy and the normalization, respectively.
Solving this variational problem leads to the $q$-canonical distribution:
\begin{equation}
p_i = \frac{1}{Z_q} \left[ 1 - (1-q)\beta E_i \right]^{\frac{1}{1-q}},
\label{eq:q_canonical}
\end{equation}
where $Z_q$ is the generalized partition function.
Furthermore, as we will see in Section \ref{sec:micro_foundations}, this distribution derived from the macroscopic variational principle exactly coincides with the effective macroscopic distribution obtained from the microscopic Superstatistics of a finite reservoir.


\section{PHYSICAL INTERPRETATION: VARENTROPY AND FINITE HEAT CAPACITY}
\label{sec:physical_interpretation}

In the previous section, we established the mathematical framework based on the $q$-logarithm and the generalized thermodynamic limit.
In this section, we show that $q$ acts as a conjugate variable to the higher-order statistics of information, specifically related to the \textit{Varentropy} and the finite heat capacity of the reservoir.

\subsection{Expansion of Renormalized Entropy and Varentropy}

To clarify the physical role of $q$, we analyze the behavior of the renormalized entropy $s_{2-q}$ in the vicinity of the standard limit $q \to 1$.
The $q$-logarithm function $\ln_q x = (x^{1-q}-1)/(1-q)$ can be Taylor expanded around $q=1$ as:
\begin{equation}
\ln_q x = \ln x + \frac{1-q}{2}(\ln x)^2 + O((1-q)^2).
\end{equation}
By substituting $x = 1/p_i$ into this equation and taking the expectation value on both sides, we obtain the exact expansion:
\begin{equation}
\begin{split}
\underbrace{\sum p_i \ln_q \frac{1}{p_i}}_{S_q(P): \text{Tsallis}} 
&= \underbrace{\sum p_i \ln \frac{1}{p_i}}_{S_{BG}(P): \text{Boltzmann-Gibbs}} \\
&\quad + \frac{1-q}{2} \left[ \underbrace{\sum p_i \left( \ln \frac{1}{p_i} - S_{BG}(P) \right)^2}_{V(P): \text{Varentropy}} + (S_{BG}(P))^2 \right] + O\left((1-q)^2\right).
\end{split}
\label{eq:entropy_expansion}
\end{equation}
Truncating the higher-order terms and writing the relation more concisely in terms of expectation values using the duality identity $\ln_q(1/x) = -\ln_{2-q}(x)$, we have the following approximation:
\begin{equation}
\underbrace{\langle -\ln_{2-q} p_i \rangle}_{S_q: \text{Tsallis}} \approx \underbrace{\langle -\ln p_i \rangle}_{S_{BG}: \text{Boltzmann-Gibbs}} + \frac{1-q}{2} \left[ \underbrace{\langle (-\ln p_i - S_{BG})^2 \rangle}_{V: \text{Varentropy}} + (S_{BG})^2 \right].
\label{eq:concise_expansion}
\end{equation}

Equation (\ref{eq:concise_expansion}) is a perturbative expansion valid in the quasi-asymptotic regime $|q-1| \ll 1$ (i.e., for small deviations from the extensive Boltzmann-Gibbs limit).
It separates the leading-order physical perturbation into the exact variance of microscopic information, known as the \textit{Varentropy}:
\begin{equation}
V(P) := \langle (-\ln p_i - S_{BG})^2 \rangle = \langle (-\ln p_i)^2 \rangle - (S_{BG})^2,
\label{eq:varentropy_def}
\end{equation}
and the square of the macroscopic background $(S_{BG})^2$.
The $(S_{BG})^2$ term appears as an algebraic consequence of isolating the exact variance $V(P)$ from the second moment of information.
While $(S_{BG})^2$ represents the global macroscopic background of the standard reference state, it is the Varentropy $V(P)$ that captures the local microscopic fluctuations.

However, this quadratic approximation has mathematical limitations. As the system moves into the strongly non-extensive regime ($1 < q < 2$), higher-order fluctuations of information (e.g., skewness and kurtosis) become non-negligible, and the truncation at the second order breaks down.

This breakdown provides the mathematical justification for adopting the full non-extensive framework rather than a truncated expansion. Although higher-order information fluctuations (skewness, kurtosis, etc.) become non-negligible, they are not independent free parameters. Within the algebraic structure of the $q$-generalized framework, this infinite tower of fluctuations is parameterized and uniquely determined by a single macroscopic scale: the non-extensivity parameter $|1-q|$, which scales the leading-order variance (Varentropy). The renormalized entropy $s_{2-q}$, governed by the $q$-logarithm, acts as the exact, non-perturbative mathematical structure that effectively \textit{resums} this single-parameter family of fluctuations into a closed function. As we will demonstrate in the next subsection and Section \ref{sec:micro_foundations}, this single mathematical scale $|1-q|$ physically originates from the finite heat capacity of the thermal environment.

Because these higher-order terms are subordinate to the variance, they do not alter the physical direction of the deformation even in the strongly non-extensive regime ($1 < q < 2$). The sign of the leading-order coefficient $(1-q)/2$ determines how the macroscopic equilibrium is driven by these information fluctuations:

\begin{itemize}
    \item \textit{Case $q=1$ (Standard Limit):} The fluctuation weight vanishes.
    The equilibrium state is determined by the mean information, leading to the standard exponential family (Boltzmann-Gibbs).
    \item \textit{Case $0 < q < 1$ (Fluctuation Prophilic):} The leading coefficient is positive.
    To maximize $S_q$, the system favors states with \textit{large} Varentropy.
    To achieve this large variance, the system heavily weights extreme values of microscopic information.
    Mathematically, this is realized by driving the probability to zero at a finite threshold, resulting in distributions with compact support where the information ($-\ln p_i$) diverges at the boundaries.
    \item \textit{Case $1 < q < 2$ (Fluctuation Phobic):} The leading coefficient is negative.
    To maximize $S_q$, the system \textit{minimizes} the Varentropy.
    In standard exponential distributions, tail states possess exponentially small probabilities, causing their information content ($-\ln p_i \propto E$) to grow linearly and contribute heavily to the variance.
    To suppress these extreme information values, the system forces the probability to decay much slower.
    Mathematically, this manifests as a heavy-tailed (power-law) distribution where the information grows logarithmically ($-\ln p_i \propto \ln E$) rather than linearly, thereby compressing the fluctuations and avoiding the Varentropy penalty inherent in exponential cutoffs.
\end{itemize}

Thus, the deviation $1-q$ acts as a macroscopic control parameter that explicitly governs the statistical weight of the microscopic information fluctuations, with the Varentropy serving as the leading-order macroscopic observable of this effect.

\subsection{Connection to Thermodynamic Heat Capacity}

In statistical mechanics, macroscopic fluctuations in entropy are linked to the heat capacity of the system.
According to the Einstein fluctuation theory~\cite{Einstein1910, Landau1980}, the fluctuation $\delta S$ must be defined carefully. Since entropy is inherently a global property of the probability distribution, there is no entropy for a single microstate. Instead, $\delta S$ represents the macroscopic manifestation of the variance in the microscopic energy exchange between the system and the finite heat bath. The corresponding microscopic observable is the energy fluctuation of the reservoir, characterized by the variance $\langle(\delta E)^2\rangle = k_B T^2 C$, where $C$ is the heat capacity of the system. This microscopic energy variance maps directly to the macroscopic entropy variance via the standard thermodynamic fluctuation formalism, yielding a Gaussian probability distribution for $\delta S$ around equilibrium:
\begin{equation}
P(\delta S) \propto \exp \left( -\frac{(\delta S)^2}{2k_B C} \right).
\end{equation}
This formula implies that the macroscopic variance of entropy is directly proportional to the heat capacity: $\langle (\delta S)^2 \rangle = k_B C$.

In our microscopic framework, this variance of information is captured by the Varentropy $V(P)$.

As established in the previous subsection, the coefficient $|1-q|$ serves as the single leading parameter that governs the scale of these information fluctuations. Comparing this variance-controlling parameter with the thermodynamic stability condition of the Einstein theory shows that $|1-q|$ acts as the generalized macroscopic cost of these fluctuations, corresponding exactly to the inverse heat capacity.
Physically, the standard limit $q \to 1$ corresponds to an infinite heat bath ($C \to \infty$), where fluctuations vanish relative to the system size.
Conversely, a deviation from unity ($q \neq 1$) implies a finite heat capacity, where the reservoir cannot fully absorb energy fluctuations.
This interpretation aligns with the result obtained by Wilk and W{\l}odarczyk~\cite{Wilk2000}, leading to the fundamental correspondence:
\begin{equation}
|1-q| \simeq \frac{1}{C}.
\label{eq:q_C_relation}
\end{equation}

This relation provides a physical interpretation of $q$ in terms of the thermal environment:
\begin{enumerate}
    \item \textit{Infinite Heat Bath ($q \to 1$):}
    When the system is coupled to an infinite reservoir ($C \to \infty$), the parameter $q$ approaches unity.
    The temperature is fixed, and the canonical ensemble is exact.
    \item \textit{Finite Heat Bath ($q \neq 1$):}
    When the reservoir is finite ($C < \infty$), the exchange of energy induces non-negligible fluctuations in the intensive parameters (temperature).
    In this regime, $q$ deviates from unity to account for the finite-size effects of the environment.
\end{enumerate}

Therefore, the non-extensive parameter $q$ is a physical measure of the \textit{finite heat capacity} of the heat bath.
The anomalous scaling $N^{2-q}$ derived in Section \ref{sec:theoretical_framework} can be interpreted as a consequence of this finite-size effect, where global temperature fluctuations of the finite reservoir induce strong correlations among the system's components, effectively reducing their independent degrees of freedom.
This mechanism provides a macroscopic formulation for a fluctuating environment, which we will microscopically demonstrate in the next section via Superstatistics.


\section{Microscopic Foundations: Superstatistics and Fluctuations}
\label{sec:micro_foundations}

\subsection{Probabilistic Inconsistency of Type-A Superstatistics}
In the original formulation of superstatistics, often referred to as Type-A superstatistics \cite{BeckCohen03}, the marginal probability distribution $P(E)$ of the energy $E$ is expressed by integrating the standard Boltzmann factor weighted by a long-term temperature fluctuation distribution $f(\beta)$:
\begin{equation}
P(E) = \int_0^\infty f(\beta) \frac{e^{-\beta E}}{Z_0} d\beta,
\end{equation}
where $Z_0$ is treated as a structural constant. However, as critically evaluated by Sattin \cite{Sattin2006} and extended by Davis \cite{Davis2022}, this configuration introduces a probabilistic inconsistency when applied to rigorous thermodynamic systems. 

If both $P(E)$ and $f(\beta)$ are strictly defined as normalized probability distributions, the partition function cannot be assumed to be a constant $Z_0$ independent of the fluctuating parameter $\beta$. Proper statistical normalization requires the explicit introduction of the $\beta$-dependent partition function $Z(\beta)$, which inherently reflects the microcanonical density of states $\Omega(E)$ of the system under consideration. Neglecting this functional dependency leads to an incomplete relation between the fluctuation of $\beta$ and the entropic index $q$, undermining the self-consistency of the macroscopic thermodynamic relations.

\subsection{Rigorous Formulation via Type-B Superstatistics}
To resolve the structural defect of the Type-A framework and to maintain a self-consistent thermodynamic formalism without invoking artificial escort distributions, it is necessary to employ the framework known as Type-B superstatistics. Within this framework, the exact conditional probability density $P(E|\beta)$ for a given inverse temperature $\beta$ is defined as:
\begin{equation}
P(E|\beta) = \frac{\Omega(E) e^{-\beta E}}{Z(\beta)},
\end{equation}
where the fundamental partition function $Z(\beta)$ is explicitly evaluated through the integration over the entire state space:
\begin{equation}
Z(\beta) = \int_0^\infty \Omega(E) e^{-\beta E} dE.
\end{equation}

Here, we consider a system characterized by a constant microcanonical heat capacity $C_{\text{sys}} = (\alpha + 1) k_B$, which yields a power-law density of states $\Omega(E) \propto E^\alpha$ (with $\alpha > -1$). Under this physical condition, the corresponding partition function is analytically evaluated as:
\begin{equation}
Z(\beta) = \frac{\Gamma(\alpha+1)}{\beta^{\alpha+1}}.
\end{equation}
Consequently, the normalized conditional probability density is expressed in the following closed form:
\begin{equation}
P(E|\beta) = \frac{\beta^{\alpha+1}}{\Gamma(\alpha+1)} E^\alpha e^{-\beta E}.
\end{equation}

\subsection{Deductive Derivation of the Variance Relation via Uniqueness of the Inverse Laplace Transform}
To preserve the deductive architecture of our formalism, we do not phenomenologically assume the functional form of the temperature fluctuation $f(\beta)$ a priori. Instead, we deduce it uniquely from the macroscopic $q$-canonical distribution established in the previous sections.

From the foundational differential equation and the Varentropy structure, the macroscopic marginal probability distribution of the system is governed by the $q$-canonical form without escort distributions:
\begin{equation}
P(E) \propto \Omega(E) \left[1 + (q-1)\beta_0 E\right]^{-\frac{1}{q-1}} \propto E^\alpha \left[1 + (q-1)\beta_0 E\right]^{-\frac{1}{q-1}}.
\label{eq:macro_q_dist}
\end{equation}
Simultaneously, within the Type-B superstatistics framework, this macroscopic distribution must be represented as the exact marginalization of the properly normalized conditional probability $P(E|\beta)$ over the unknown temperature fluctuation distribution $f(\beta)$:
\begin{equation}
P(E) = \int_0^\infty f(\beta) \frac{\beta^{\alpha+1}}{\Gamma(\alpha+1)} E^\alpha e^{-\beta E} d\beta.
\label{eq:marginal_type_b}
\end{equation}

Equating Eq.\ (\ref{eq:macro_q_dist}) and Eq.\ (\ref{eq:marginal_type_b}), and factoring out the purely density-of-states-dependent term $E^\alpha$, we obtain the following integral equation:
\begin{equation}
\left[1 + (q-1)\beta_0 E\right]^{-\frac{1}{q-1}} \propto \int_0^\infty \left[ f(\beta) \beta^{\alpha+1} \right] e^{-\beta E} d\beta.
\end{equation}
The right-hand side is exactly the Laplace transform $\mathcal{L}$ of the modified function $g(\beta) := f(\beta) \beta^{\alpha+1}$.
The equation can be rewritten as:
\begin{equation}
\mathcal{L}\{f(\beta) \beta^{\alpha+1}\}(E) \propto \left[1 + (q-1)\beta_0 E\right]^{-\frac{1}{q-1}}.
\end{equation}

Due to the fundamental theorem of the \textit{uniqueness of the inverse Laplace transform}, the function inside the transform is uniquely determined. Applying the standard inverse Laplace transform $\mathcal{L}^{-1}\{(1+aE)^{-\nu}\} \propto \beta^{\nu-1} e^{-\beta/a}$, we obtain:
\begin{equation}
f(\beta) \beta^{\alpha+1} \propto \beta^{\frac{1}{q-1}-1} e^{-\frac{\beta}{(q-1)\beta_0}}.
\end{equation}
Solving for the fluctuation distribution $f(\beta)$, we find:
\begin{equation}
f(\beta) \propto \beta^{\left(\frac{1}{q-1} - \alpha - 1\right) - 1} e^{-\frac{\beta}{(q-1)\beta_0}}.
\end{equation}

This mathematical inversion proves that $f(\beta)$ is uniquely a Gamma distribution $f(\beta) \propto \beta^{c-1} e^{-\beta/b}$ with the scale parameter $b = (q-1)\beta_0$ and the constrained shape parameter:
\begin{equation}
c = \frac{1}{q-1} - (\alpha + 1) = \frac{1 + (1-q)(\alpha+1)}{q-1}.
\label{eq:shape_parameter_c}
\end{equation}

Since the relative variance of a standard Gamma distribution is given by $\sigma_\beta^2 / \langle\beta\rangle^2 = 1/c$, substituting the unique shape parameter $c$ directly yields the exact variance relation:
\begin{equation}
\frac{\sigma_\beta^2}{\langle\beta\rangle^2} = \frac{q-1}{1 + (1-q)(\alpha + 1)}.
\label{eq:new_variance_relation}
\end{equation}

This completes the deductive derivation. We have demonstrated that the exact relation identified by Davis \cite{Davis2022_classification} is not merely a phenomenological observation but a necessary consequence mathematically enforced by the uniqueness of the inverse Laplace transform, connecting the microcanonical density of states, the Type-B probability normalization, and the fundamental Varentropy structure.

\subsection{Physical Identification: Heat Bath Degrees of Freedom and the Thermodynamic Limit}
The mathematical inversion established above dictates that the temperature fluctuations must follow a Gamma distribution. Physically, this macroscopic fluctuation distribution represents the local temperature fluctuations arising from a sum of independent Gaussian variables (e.g., kinetic energies) constituting a finite heat bath.

We can physically identify the shape parameter $c$ of the deduced Gamma distribution with half the effective degrees of freedom of the heat bath, denoted by $n/2$. Under this physical identification, the distribution $f(\beta)$ explicitly corresponds to a scaled $\chi^2$ distribution. To ensure that the mean inverse temperature remains fixed at $\langle\beta\rangle = \beta_0$ regardless of the bath size $n$, the scale parameter $b$ must scale as $b = \langle\beta\rangle / c = 2\beta_0 / n$. Consequently, the variance of the inverse temperature is given by $\sigma_\beta^2 = c b^2 = 2\beta_0^2 / n$. This yields the relative variance of the heat bath's temperature fluctuations:
\begin{equation}
\frac{\sigma_\beta^2}{\langle\beta\rangle^2} = \frac{2}{n}.
\label{eq:physical_relative_variance}
\end{equation}

By equating this physical relative variance associated with the heat bath [Eq.\ (\ref{eq:physical_relative_variance})] with the exact mathematical constraint derived from Type-B superstatistics [Eq.\ (\ref{eq:new_variance_relation})], we establish the fundamental relation connecting the non-extensivity parameter $q$, the system's microcanonical state density exponent $\alpha$, and the heat bath's degrees of freedom $n$:
\begin{equation}
\frac{2}{n} = \frac{q-1}{1 + (1-q)(\alpha+1)}.
\end{equation}
Solving this relation explicitly for $q$, we obtain:
\begin{equation}
q = 1 + \frac{2}{n + 2(\alpha+1)}.
\label{eq:exact_q_n_alpha}
\end{equation}

Equation (\ref{eq:exact_q_n_alpha}) is an analytical result of our deductive framework. It demonstrates that the non-extensive deformation $q \neq 1$ is jointly determined by the finite size of the heat bath ($n$) and the intrinsic heat capacity of the system itself parameterized by $\alpha$. The effective total degrees of freedom governing the equilibrium is the exact sum $n_{\text{eff}} = n + 2(\alpha+1)$. When the system's internal degrees of freedom are negligibly small compared to the macroscopic heat bath ($\alpha \ll n$), we recover the standard phenomenological approximation $q \simeq 1 + 2/n$.

\begin{figure}[htbp]
\centering
\includegraphics[width=0.8\linewidth]{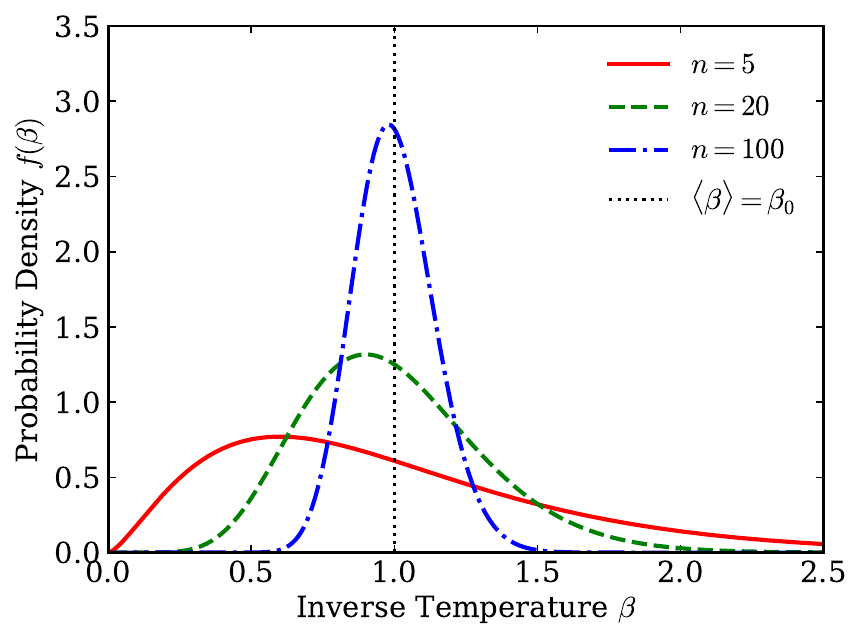}
\caption{Microscopic temperature fluctuations described by the Gamma distribution $f(\beta)$ for varying degrees of freedom $n$ of the heat bath. The mean inverse temperature is fixed at $\langle \beta \rangle = \beta_0$. As $n$ increases ($n \to \infty$), the distribution systematically narrows and converges to a Dirac delta function $\delta(\beta - \beta_0)$, physically corresponding to an exact canonical ensemble coupled to an infinite heat bath.}
\label{fig:gamma_fluctuations}
\end{figure}

Finally, in the thermodynamic limit of an infinite heat bath ($n \to \infty$), the relative variance systematically vanishes. As illustrated in Fig.~\ref{fig:gamma_fluctuations}, the Gamma distribution $f(\beta)$ systematically narrows and converges to the Dirac delta function $\delta(\beta - \beta_0)$. In this infinite limit, the superposition integral reduces to the standard Boltzmann factor ($P(E) \to e^{-\beta_0 E}$), ensuring that $q \to 1$ and the classical Boltzmann-Gibbs statistics are exactly recovered.

\paragraph{Remark on Generalized Logarithms and Physical Consistency}
Expanding alternative generalized logarithms, such as the $\phi$-logarithm introduced by Naudts \cite{Naudts2011}, mathematically generates a variance-like term in the second-order expansion of the corresponding entropic functions. However, obtaining a mathematical variance through Taylor expansion is distinct from endowing the resulting structure with a self-consistent thermodynamic interpretation. 

Within the framework of the $\phi$-logarithm, defined by $\ln_\phi(p) := \int_{1}^{p} \frac{1}{\phi(v)} dv$, numerous generalized distributions can be formally constructed. However, imposing the strict physical requirement that the undeformed, original Varentropy (the intrinsic microscopic variance of information) must macroscopically manifest as the natural temperature fluctuations of a finite heat bath---represented by the Gamma distribution in the Type-B superstatistics detailed above---uniquely constrains the underlying geometry of the state space. It is only when $\phi(p) = p^q$ (the Tsallis $q$-logarithm) that this physically realizable Gamma distribution is exactly and self-consistently recovered without relying on artificial escort distributions or phenomenological drift terms.
Other choices of $\phi(p)$ fail to satisfy the simultaneous requirements of finite heat capacity, physical fluctuation models, and macro-micro thermodynamic stability established in this work.


\section{Macroscopic Consequences: Heat Capacity and Anomalous Scaling}
\label{sec:macroscopic_consequences}

\subsection{Thermodynamic Stability and Heat Capacity}
In standard statistical mechanics, the temperature fluctuations of a system coupled with an environment are derived directly from the energy fluctuation $\langle(\delta E)^2\rangle = k_B T^2 C$. Using the thermodynamic relation $\delta E = C \delta T$, this leads to the Landau fluctuation formula for the relative variance of temperature:
\begin{equation}
\frac{\sigma_T^2}{T^2} = \frac{k_B}{C},
\label{eq:landau_fluctuation}
\end{equation}
where $C$ is the dimensionless heat capacity in natural units ($k_B=1$). Since the inverse temperature is defined as $\beta = 1/T$, a first-order expansion of $\beta$ around the mean temperature $\langle T \rangle$ yields $\delta \beta \approx - \langle \beta \rangle \delta T / \langle T \rangle$. Squaring and taking the expectation value leads to the equivalence of the relative variances:
\begin{equation}
\frac{\sigma_\beta^2}{\langle \beta \rangle^2} \approx \frac{\sigma_T^2}{\langle T \rangle^2} = \frac{1}{C}.
\label{eq:relative_variance_landau}
\end{equation}
This equivalence demonstrates that the relative variance of the inverse temperature $\beta$ perfectly aligns with the standard macroscopic temperature fluctuation $\sigma_T^2 / \langle T \rangle^2 = 1/C$ derived from Landau's fluctuation theory. To explicitly connect this macroscopic thermodynamic result to the non-extensive parameter $q$, we must evaluate the total heat capacity $C$ of the finite system.

Within our framework, the total dimensionless heat capacity of the combined system and environment can be expressed as the sum of the heat capacity of the heat bath, $C_{\text{bath}} = n/2$, and the microcanonical heat capacity of the system itself, $C_{\text{sys}} = \alpha + 1$. Thus, the total heat capacity is given by:
\begin{equation}
C = C_{\text{bath}} + C_{\text{sys}} = \frac{n}{2} + (\alpha + 1).
\label{eq:total_heat_capacity}
\end{equation}

By substituting this total heat capacity [Eq.\ (\ref{eq:total_heat_capacity})] into the relative variance relation [Eq.\ (\ref{eq:relative_variance_landau})], the fluctuation scale is physically quantified as:
\begin{equation}
\frac{\sigma_\beta^2}{\langle \beta \rangle^2} = \frac{1}{\frac{n}{2} + (\alpha + 1)} = \frac{2}{n + 2(\alpha+1)}.
\label{eq:macro_relative_variance}
\end{equation}
Comparing this macroscopic result [Eq.\ (\ref{eq:macro_relative_variance})] with the exact structural relation derived earlier from the microscopic Type-B framework [Eq.\ (\ref{eq:exact_q_n_alpha})], we firmly establish the fundamental identity:
\begin{equation}
q - 1 = \frac{1}{C}.
\label{eq:q_C_relation_final}
\end{equation}
Thus, the non-extensive parameter $q$ is proven to be the exact macroscopic manifestation of the finite heat capacity $C$ of the reservoir.
As a specific physical realization, the exact phase-space integration for a finite monatomic ideal gas also naturally yields the consistent relation $q - 1 = 1/C$ \cite{Lima2020}.

\begin{figure}[htbp]
\centering
\includegraphics[width=0.8\linewidth]{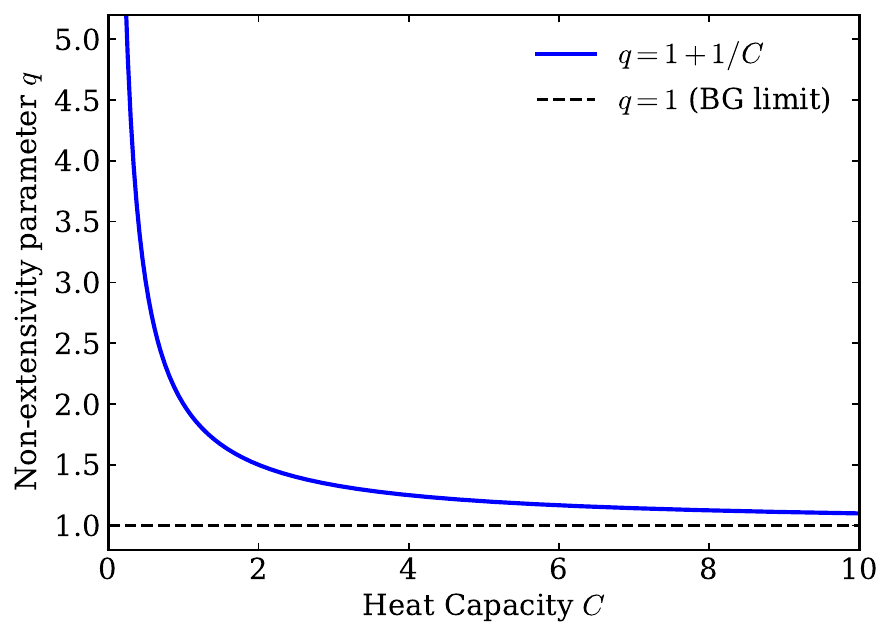}
\caption{Relationship between the non-extensivity parameter $q$ and the dimensionless heat capacity $C$ of the reservoir. As the heat capacity diverges in the standard thermodynamic limit ($C \to \infty$), the fluctuations vanish and the system recovers Boltzmann-Gibbs statistics ($q \to 1$). For finite systems with bounded $C$, the parameter $q > 1$ acts as a thermodynamic stabilizer absorbing the macroscopic fluctuations.}
\label{fig:q_vs_C}
\end{figure}

Equation (\ref{eq:q_C_relation_final}) establishes that the non-extensivity parameter deviation $q-1$ is the reciprocal of the total thermodynamic heat capacity. In the standard thermodynamic limit where the reservoir becomes infinite ($C \to \infty$), the parameter $q$ continuously approaches unity, thereby recovering the Boltzmann-Gibbs limit. For finite systems where $C$ remains bounded, the parameter $q > 1$ acts as a thermodynamic stabilizer that absorbs global fluctuations into the state definitions, preventing thermodynamic singularities (see Fig.~\ref{fig:q_vs_C}).

\subsection{Anomalous Scaling}
The fundamental relation $q - 1 = 1/C$ has implications for the thermodynamic scaling properties of correlated systems. If the parameter $q$ remains strictly bounded away from unity ($q > 1$) as the system size $N \to \infty$, Eq.\ (\ref{eq:q_C_relation_final}) dictates that the total heat capacity $C(N)$ must saturate to a finite, constant value:
\begin{equation}
C(N) = \frac{1}{q-1} \sim \text{const}.
\end{equation}
This behavior characterizes a strongly non-extensive regime where the specific heat capacity vanishes in the large system limit ($C(N)/N \to 0$). Such thermodynamic saturation is consistent with systems constrained by global long-range correlations, where standard linear extensivity is structurally violated \cite{Campa2009}, directly addressing the long-standing skepticism regarding the thermodynamic consistency of such ensembles \cite{Lutsko2011}.

Furthermore, for physical systems obeying area laws, such as black hole thermodynamics or strongly entangled quantum systems, the total entropy scales non-extensively as $S \propto N^{\gamma}$ with $\gamma < 1$ \cite{Tsallis2013}. In the language of our framework, this corresponds to an effective thermodynamic heat capacity scaling sub-linearly as $C(N) \propto N^{\gamma}$. Consequently, the entropic parameter $q$ approaches the standard Boltzmann-Gibbs limit asymptotically as a power law:
\begin{equation}
q - 1 \sim N^{-\gamma}.
\end{equation}
This mathematical correspondence demonstrates that the derived relation $q-1 = 1/C$ provides a first-principles thermodynamic foundation for various anomalous scaling laws observed in complex physical systems, resolving the long-standing stability debate without relying on phenomenological adjustments.


\section{CONCLUSION AND PERSPECTIVES}
\label{sec:conclusion}

In this paper, we have established a consistent thermodynamic framework for non-extensive systems, grounded in the scaling properties of the $q$-factorial and the physical concept of Varentropy. We addressed the thermodynamic limit problem for power-law statistics by introducing the renormalized entropy $s_{2-q}$. Unlike the original Tsallis entropy, where extensivity is recovered only for a specific unique index $q$ depending on the system's correlations, the renormalized entropy $s_{2-q}$ is constructed to remain finite ($O(N^0)$) in the thermodynamic limit for an arbitrary parameter $q$, serving as the correct intensive state variable for systems with strong correlations.

\paragraph*{The Physical Meaning of $q$.---}
Through the ``Trinity of Varentropy''---Finiteness, Fluctuations, and Stability---we have clarified the physical origin of the non-extensivity parameter $q$ via a deductive construction. We demonstrated that the structural deviation $q - 1$ is the reciprocal of the total thermodynamic heat capacity $C$ of the combined system and environment:
\begin{equation}
q - 1 = \frac{1}{C}.
\end{equation}
This exact identity is deductively established without relying on phenomenological adjustments or artificial escort distributions. While prior generalized statistical frameworks have classified temperature fluctuations from a macroscopic or informational standpoint \cite{Davis2022_classification}, the present framework systematically deduces the macroscopic thermodynamic structure from the foundational microscopic differential equation governing the phase space growth. Within this deductive sequence, the uniqueness of the inverse Laplace transform mathematically enforces that the functional form of these fluctuations is uniquely restricted to a Gamma distribution. Consequently, standard Boltzmann-Gibbs statistics ($q=1$) emerges exclusively in the limit of an infinite heat bath ($C \to \infty$), whereas for finite systems with bounded resources, the microscopic variance of information (Varentropy) generated from the underlying growth rule necessarily dictates the $q$-generalized description to maintain macroscopic stability.

\paragraph*{Origin in Information Theory.---}
The concept of Varentropy originates in Information Theory, specifically in the analysis of \textit{finite block-length coding}~\cite{Strassen62, PPV10}. Just as the variance of information becomes non-negligible when the data block length is finite, the thermodynamic fluctuations become significant when the heat capacity is finite. Our result $q-1 = 1/C$ establishes an isomorphism: Tsallis statistics is to Boltzmann-Gibbs thermodynamics what finite-length coding is to asymptotic Shannon theory. Both frameworks represent the generalization needed to describe systems constrained by \textit{finite resources (i.e., finite block-length and finite heat capacity)}. This thermodynamic perspective is mathematically complemented by recent developments in generalized limit theorems \cite{PartI} and the rigorous algebraic formulation of finite-size effects in complex systems \cite{PartVI}.

\paragraph*{Future Outlook.---}
The framework presented here provides a theoretical basis for applying non-extensive statistics to small-scale systems where fluctuations are inherent. Applications include nanothermodynamics, the statistics of single-molecule experiments, and complex networks where the effective degrees of freedom are limited.

\section*{Acknowledgements}
This work was supported by JSPS KAKENHI Grant Number 26K14703. The author thanks the anonymous reviewers for their constructive comments and valuable suggestions, which significantly improved the mathematical rigor and physical clarity of this manuscript.







\bibliographystyle{elsarticle-num}
\bibliography{reference.bib} 
\end{document}